\documentclass[preprint,showpacs,aps]{revtex4}
\begin{document} 
\draft 
\preprint{UM-P-91/47}
\title{On Quantization, the Generalized Schr\"{o}dinger Equation
and Classical Mechanics\footnote{Author's note: This is the original text of a pre-print that was lost for some twenty years. Over the years, people have asked me for a copy but I had none. This work is {\em not refereed\/} and was never published (since it was lost). Given what I know now, I would write it differently, but place it here under Creative Commons 3.0 - Attribution License under the condition that this footnote is displayed. It is old, but may be of interest for the ideas it explores. Opinions expressed may have been superseded by later works.}}
\date{Original Preprint: June 1991}
\author{Kingsley R. W. Jones}
\address{School of Physics, University of Melbourne, Parkville 3052,\\
Melbourne, Australia}
\begin{abstract}
Using a new state-dependent, $\lambda$-deformable, linear functional operator, 
${\cal Q}_{\psi}^{\lambda}$, which presents a natural $C^{\infty}$ deformation
of quantization, we obtain a uniquely selected non--linear, integro--differential
Generalized Schr\"{o}dinger equation. The case ${\cal Q}_{\psi}^{1}$ reproduces
linear quantum mechanics, whereas ${\cal Q}_{\psi}^{0}$ admits an exact dynamic,
energetic and measurement theoretic {\em reproduction\/} of classical mechanics.
All solutions to the resulting classical wave equation are given and we show that
functionally chaotic dynamics exists.    
\end{abstract}  
\pacs{0230, 0365, 0545}
\maketitle

This paper reports some new and quite elementary results that we believe must
lie at the very heart of the connection between classical and quantum
mechanics. In the interests of rapid communication we shall not describe how
we obtained them. This omission will be rectified in subsequent publications.

Our work appears in the spirit of de Broglie\cite{brog}. It is inspired
by the early work of Dirac\cite{dirac1,dirac2}, but evolved purely from a
single geometric picture\cite{jones1} that grew out of our parallel measurement
theoretic research\cite{jones2}, coupled with thinking stimulated by a single
prescient paper by Klauder\cite{klau}. The work potentially opens up a vast new
territory of quantal dynamical systems research. 

Perhaps the most topical result we shall demonstrate is that quantum chaos
{\em certainly\/} exists within a dynamical system that contains standard
quantum mechanics. Moreover, the nonlinear dynamics described appears to fit
within the recently elaborated formal scheme of Weinberg\cite{wein}. However,
this work does more than that. It both solves an old problem and generates a
new dynamical structure of obvious physical application. This structure should
prove useful both for phenomenological model building, and as a theoretical
guide to some potentially exciting new physics at the boundary of the quantal
and classical domains. 

A brief outline is in order. We first define quantization, then deformed
quantization. From the latter we derive the Generalized Schr\"{o}dinger
equation. This includes both the usual Schr\"{o}dinger equation and a new
Classical Schr\"{o}dinger equation as particular values of $\lambda$. We
exhibit the entire infinite family of solutions of the latter equation in
parametric form, and then show that it gives an exact energetic, dynamic and
measurement theoretic {\em reproduction\/} of classical mechanics via a
travelling wave double solution.

Consider any classical phase space function $f_{c}({\bf q},{\bf p})$ of 
$n$--degrees of freedom. Quantization of this phase space function amounts
to application of the following $\psi$-dependent linear functional operator:
\begin{equation}
\label{quantize}
{\cal Q}_{\psi} \equiv \exp
\left\{\sum_{k=1}^{n} 
(\hat{q}_{k} -\langle \hat{q}_{k} \rangle_{\psi})
\frac{\partial}{\partial q_{k}} +
(\hat{p}_{k} -\langle \hat{p}_{k} \rangle_{\psi})
\frac{\partial}{\partial p_{k}}\right\}.
\end{equation}
The domain of this map is any region of phase space where $f_{c}({\bf q},{\bf
p})$ is analytic, its range the set of quantal operators. The domain of the
resulting quantal operator proves to be precisely the set of all $\psi$ 
such that the numerical values of the expectation values $\langle \hat{q}_{k}
\rangle_{\psi}$ and  $\langle \hat{p}_{k} \rangle_{\psi}$ lie in an analytic
domain of the initial classical function $f_{c}({\bf q},{\bf p})$. 

The linear operator (\ref{quantize}) is to be understood as a formal power
series. It generates a natural, uniquely defined, Weyl--ordered operator 
Taylor series whose action upon any classical phase space function is:
${\cal Q}_{\psi}\circ f_{c}({\bf q},{\bf p}) = 
\hat{f}_{q}(\hat{\bf q},\hat{\bf p}),$
where $q_{k}$ and $p_{k}$ are general classical canonical coordinates
satisfying the rules: $\{q_{j},q_{k}\} =0$, $\{p_{j},p_{k}\} = 0$ and
$\{q_{j},p_{k}\} = \delta_{ij},$ whereas $\hat{q}_{k}$ and $\hat{p}_{k}$ are
general quantal canonical operators satisfying the rules
$[\hat{q}_{j},\hat{q}_{k}] =0$, $[\hat{p}_{j},\hat{p}_{k}] = 0$ and
$[\hat{q}_{j},\hat{p}_{k}] = i\hbar\delta_{ij}\hat{1}$. The classical
differential operators $\partial q_{k}$ and $\partial p_{k}$ are taken to
commute with $\hat{q}_{k}$ and $\hat{p}_{k}$ and merely evaluate derivatives
of the classical function $f({\bf q},{\bf p})$ in a $\psi$-dependent way at
the quantal expectation values $\langle\hat{q}_{k}\rangle_{\psi}$ and 
$\langle\hat{p}_{k}\rangle_{\psi}$. However, note that the $c$--number
derivatives need not be taken to commute with one another so that
non--analytic behaviour such as
$\partial^{2}_{qp}f(q,p)\ne\partial^{2}_{pq}f(q,p)$ is also catered for.

Because the $\hat{q}_{j}$ and $\hat{p}_{j}$ do not commute, the generalized
Taylor series implict in (\ref{quantize}) should not have like order terms
collected. This is the source of its important Weyl--ordering property. An
example will suffice. Consider the one dimensional generalized harmonic
oscillator:  
$ H_{c}(q,p) = \frac{1}{2}(a p^{2} + b pq + c q^{2})$. 
Applying (\ref{quantize}) we obtain
$$ \hat{H}_{q}(\hat{q},\hat{p}) \equiv
{\cal Q}_{\psi} \circ H_{c}(q,p) =
\frac{1}{2}( a\hat{p}^{2} + b/2[\hat{p}\hat{q} + \hat{q}\hat{p}]
           + c\hat{q}^{2}).$$
The $\psi$-dependence of the result has disappeared. This may be understood via
the deep observation of Weierstrass\cite{note} that analytic functions can be
considered as a family of power series whose totality comprises the analytic
function and whose consistency of definition requires only that the individual
members of that family should share continuously connected overlapping domains
of agreement. The construction of the full quantal operator therefore amounts
to a kind of analytic continuation in the wave function $\psi$, with reference
to the analytic domain of the starting classical function via the expectation
values of each $\psi$ so considered.

The above we believe to be a new result. It systematizes a significant body of previously known results\cite{mess}, resolves operator ordering ambiguities, and suggests new results as well\cite{jones1}. 

Let us now introduce the $\lambda$--deformed operator:  
\begin{equation}
\label{quantize2}
{\cal Q}^{\lambda}_{\psi} \equiv \exp
\left\{\sum_{k=1}^{n} 
\lambda (\hat{q}_{k} -\langle \hat{q}_{k} \rangle_{\psi})
\frac{\partial}{\partial q_{k}} +
\lambda (\hat{p}_{k} -\langle \hat{p}_{k} \rangle_{\psi})
\frac{\partial}{\partial p_{k}}\right\}.
\end{equation}
This is to be understood in exactly the same manner as (\ref{quantize}). 
However, for $\lambda\ne 1$ the properties of (\ref{quantize}) as a Taylor
series are modified. In general, (\ref{quantize2}) produces a 
$\psi$-dependent quantization prescription:
$${\cal Q}_{\psi}^{\lambda}:f_{c}({\bf q},{\bf p}) \mapsto
\hat{f}_{q}(\psi;\lambda) =  \hat{f}_{q}^{\lambda}(
\langle\hat{\bf q}\rangle_{\psi} +\lambda  (\hat{\bf q}-\langle\hat{\bf q}\rangle_{\psi}),
\langle\hat{\bf p}\rangle_{\psi} +\lambda  (\hat{\bf p}-\langle\hat{\bf p}\rangle_{\psi})) .$$
Moreover, it is clear that this happens in a smooth fashion and yields an
entirely natural and inifinitely differentiable deformation of the dynamical
structure of quantum mechanics. 

Application of the deformed rule (\ref{quantize2}) now produces a new family
of non--linear integro--differential equations, which family we elect to call
{\em The Generalized Schr\"{o}dinger equation\/}. 

To $H_{c}({\bf q},{\bf p})$ we apply ${\cal Q}_{\psi}^{\lambda}$ to obtain   
$\hat{H}_{q}(\psi;\lambda) \equiv 
{\cal Q}_{\psi}^{\lambda}\circ H_{c}({\bf q},{\bf p}),$
This yields its Generalized Schr\"{o}dinger equation as
\begin{equation}
\label{general}
i\hbar \frac{d}{dt}|\psi\rangle = \hat{H}_{q}(\psi;\lambda)|\psi\rangle.
\end{equation}
A precursorial formal structure into which (\ref{general}) appears to fall
has been given recently by Weinberg\cite{wein}.

Significantly, $\lambda$ is a mathematically free parameter, though if this
is new physics then its value in any physical situation probably includes
$\hbar$ in dimensionless combination with other parameters.

The family (\ref{general}) incorporates the following physically plausible
features: a state dependent energy behaviour in the non--linear regime
($\lambda\ne1$); unitary  but non distance preserving dynamics; true quantum
chaos and representation independent physical predictions. As noted by
numerous authors\cite{wein,bial} the final condition is vital to the survival
of symmetry as a physical principle. 

Note that the $\psi$--dependence of $\hat{H}_{q}(\psi;\lambda)$ enters in this
case purely via the representation independent quantities:
$\langle{\bf\hat{q}}\rangle_{\psi}$  and
$\langle{\bf\hat{p}}\rangle_{\psi}$. 
This, coupled with the fact that the dynamical generator is Hermitian, ensures
that we have a $\psi$--dynamics that is at all times unitary and norm
preserving, but one that need not preserve the natural quantal metric
distance\cite{prov}: $d(\psi_{1},\psi_{2})\equiv   1 - |\langle
\psi_{1}|\psi_{2}\rangle|^{2}$. It follows that the superposition
principle is no longer valid in the regime $\lambda\ne 1$. This may well
require careful consideration when attempting to construct a dynamics for
non--separable systems\cite{bial}.

True quantum chaos is now indicated as the sensitive dependence of
wavefunction trajectories upon wavefunction initial conditions\cite{note3}.
We now exhibit such chaotic wavefunction dynamics for ${\cal Q}_{\psi}^{0}$
quantised classically chaotic dynamics. We call the $\lambda =0$ case the
Classical Schr\"{o}dinger equation.

In coordinate representation the Classical Schr\"{o}dinger equation corresponds
to a nonlinear first order integro--differential equation. For the one
dimensional classical Hamiltonian:  
$$H_{c}(q,p) = \frac{p^{2}}{2m} + V(q),$$ 
It has the explicit form 
\begin{equation}
\label{class}
i\hbar \left(\frac{\partial }{\partial t}
+            \frac{{\langle p \rangle}}{m} 
             \frac{\partial }{\partial q}\right) \psi(q,t) = 
\left (V(\langle q \rangle) 
+ \frac{\partial V}{\partial q}(\langle q \rangle)  
(q - \langle q \rangle)  
-\frac{\langle p \rangle^{2}}{2m} \right) \psi(q,t),
\end{equation}
where of course $\langle q \rangle$ and  $\langle p \rangle$ must at all times
satisfy the relations
$$ \langle q \rangle = \int_{-\infty}^{\infty}\!
q\psi^{*}(q,t)\psi(q,t)\,dq
\;\; \mbox{and} \;\;
\langle p \rangle = \int_{-\infty}^{\infty}\!
-i\hbar\psi^{*}(q,t)\frac{\partial}{\partial q}\psi(q,t)\,dq.$$
Such an equation is rather difficult to solve in general. Here are all of its
solutions:
\begin{equation}
\label{answer}
\psi(q,t) =  e^{i\phi(t)} 
\exp\left\{-\frac{i}{2\hbar} \langle p \rangle\langle q \rangle \right\}
\exp\left\{+ \frac{i}{\hbar} \langle p \rangle  q \right\} 
\psi_{0}(q - \langle q \rangle),
\end{equation}
where $\psi_{0}(q)$ belongs to the infinite family of square integrable and
differentiable wavefunctions whose expectation values for both position and
momentum operators are equal to zero. The phase factor is
\begin{equation}
\label{phase}
\phi(t) = \frac{1}{\hbar}\int_{0}^{t}
\left[\frac{1}{2}\left(
\langle p \rangle \frac{d}{dt} \langle q \rangle -
\langle q \rangle \frac{d}{dt} \langle p \rangle \right)
- \frac{\langle p \rangle^{2}}{2m} - V(\langle q \rangle)
\right]\,d\tau.
\end{equation}
To explain how (\ref{answer}) provides an infinite family of solutions we must
point out that the terms $\langle q \rangle$ and $\langle p \rangle$ are time
dependent parameters that enforce the functional evolution of $\psi(q,t)$.
The required values of these parameters are obtained at all times by solving 
the purely classical equations:
\begin{equation}
\label{ham} 
\frac{d \langle q \rangle}{dt} = 
\frac{\partial H_{c}}{\partial p}(\langle q \rangle,\langle p \rangle),
\;\;\mbox{and}\;\;
\frac{d \langle p \rangle}{dt} = 
-\frac{\partial H_{c}}{\partial q}(\langle q \rangle,\langle p \rangle).
\end{equation}
In addition, due to the special form of (\ref{answer}), these parameters will
always correspond to the required expectation values. In this way 
(\ref{class}) realizes Louis de Broglie's dream of the double
solution\cite{brog}.

In order to prove the above one simply substitutes (\ref{answer}) into
(\ref{class}). Taking real and imaginary parts of the resulting equation, one 
can then derive (\ref{ham}) as an essential consistency requirement. Many
$\psi_{0}(q)$ will do, so that (\ref{class}) actually admits an infinite 
family of non--dispersive travelling wave solutions whose expectation values
follow precisely the trajectories of the classical Hamiltonian evaluated upon
each unique trajectory associated with each one of an inifinite family of
equivalent initial conditions (choice of $\psi_{0}(q)$). This is the highly
desirable version of Ehrenfest's theorem which one cannot derive from linear
quantal mechanics\cite{mess,gott}. The proof that (\ref{answer}) generates all
of the solutions we shall present elsewhere\cite{jones1}.

We remark that the ${\cal Q}_{\psi}^{1}$ harmonic oscillator has a zero point
energy, whereas the ${\cal Q}_{\psi}^{0}$ version does not, and $\lambda$
interpolates between the two. The existence of two possible quantizations, in
this case, was commented upon a long time ago by Klauder\cite{klau}, who
appears to have been the first to encounter a manifestation of deformed
quantization.

Note that the phase (\ref{phase}) includes an Aharanov--Anandan\cite{ahar}
contribution such that the total value upon closed classical trajectories is
numerically equal to the {\em classical action\/}\cite{dirac2}. Moreover, the
geometric component of the ${\cal Q}_{\psi}^{0}$ phase corresponds precisely
to the abbreviated action upon closed classical trajectories. This injects a
natural quantal geometric phase into classical mechanics and suggests that
these phases are the natural {\em action\/} variables of integrable quantal
dynamics\cite{jones1}. 

Indeed for closed circuits $\Gamma$ of a ${\cal Q}_{\psi}^{0}$ quantised
$n$--degree of freedom classical system this geometric phase can be
expressed in terms of the first Poincar\'{e} integral invariant of classical
mechanics\cite{arn} by the simple expression
\begin{equation} 
\label{phase2}
\gamma(\Gamma) =\frac{1}{\hbar} \oint \sum_{j=1}^{n} p_{j}\,dq_{j}.
\end{equation} 
One now understands Einstein's version\cite{ein} of the {\em old\/}
Bohr--Sommerfeld quantization condition to be a constraint upon the geometric
phase ($2\pi\times${\em integer\/}). To the ${\cal Q}_{\psi}^{0}$ quantised
integrable $n$--degree of freedom classical system there are $n$ geometric
phase actions $\gamma_{j}$ associated with the natural wave function evolution
about a {\em functional torus\/} parametrised by $\langle q\rangle$ and
$\langle p \rangle$. The values of these phases are obtained from
(\ref{phase2}) via integration upon the $n$ irreducible contours of the
classical $n$--torus, which the functional parameters $\langle q\rangle$ and 
$\langle p \rangle$ explore\cite{arn}. 

Note that (\ref{class}) will exhibit regions of chaotic wavefunction dynamics 
for any classically chaotic Hamiltonian system.

Turning now to energetic and measurement theoretic considerations, one can
readily verify that ${\cal Q}_{\psi}^{0}$ quantized classical mechanics yields
precisely the correct classical energy at all times where this is encoded in
the {\em expectation\/} value form: $H_{c}(\langle q \rangle ,\langle p
\rangle,t) = \langle \psi(t)|\hat{H}_{q}(\psi(t);0)|\psi (t)\rangle$. This
quantal model of classical mechanics therefore amounts to a conservative
theory, just as one would expect, albeit one with $\psi$--dependent energy
dispersion. 

Changing $\hbar$ does not alter the dynamical behaviour of the equation
(\ref{class}). It simply rescales phase space. However, the {\em quantum
measurement theory\/} is changed under such rescaling. For instance, a simple
argument based upon the Heisenberg Uncertainty Principle  $(\Delta
x)^{2}(\Delta p)^{2}\ge\hbar^{2}/4$ shows that we might only expect to follow
features of a wavefunction trajectory that are large in comparison to a
confidence region $\Delta$ of area: $\int_{\Delta} dp\wedge dq \sim \hbar.$
More exotic arguments based upon modern quantum measurement theory yield
essentially the same conclusion\cite{art,hels}.

To conclude, ${\cal Q}_{\psi}^{\lambda}$ has the characteristic aesthetic
appeal of a most natural mathematical formalism. There is clearly considerable
scope for its extension (renormalized, deformed Taylor series may have
properties of general interest). The resulting Generalized Schr\"{o}dinger
equation provides a sharp spur towards the general development of the theory
of integro--differential equations and clearly has great potential physical
application. Perturbative studies in $\lambda$ and the investigation of
physically coupled dynamics in $\lambda$ are indicated\cite{note4}. Studies of
this kind offer a natural route to the exploration of {\em functional
attractors\/}, which would be likely to have important physical repercussions.

In this, a physical paper, we are more directly and vitally concerned with the
interpretation of $\lambda$. In view of the very natural appearance of this
dimensionless number we suggest that it encapsulates the degree to which a
quantal system may back react upon its environment. It is then entirely
plausible that the energy should be $\psi$--dependent (self--energy).

An empirical test that $\lambda$ need not be unity is needed. The author can
think of two possibilities. Classical mechanics has enjoyed a rather curious
success in explaining certain features of the microwave ionization spectroscopy
of Rydberg states in atomic hydrogen\cite{ryd1,ryd2}. One might envisage
nonlinearity entering into this phenomenon due to the very high order photon
processes involved. Another possibility is the observed nonlinear polarisation
dynamics of optical beams traversing media of significant third-order
nonlinear susceptibility\cite{shen,david}. It is suggested that $\lambda$
may enter here as an appropriate ratio of the linear and nonlinear
susceptibilities. A general search for such quantal two state nonlinear
dynamics is indicated.

Measurement theoretic tools for the detection of nonlinear quantal dynamics 
are reported in\cite{jones2}. On this note we might add that the measurement
theoretic consequences of a workable quantal nonlinearity are profound and may
lead to some unexpected results when coupled with current research into
nonlinear dynamics.

I am grateful to many people for their help, advice or comments at different
stages: Profs. Bruce McKellar and Ian Percival and Drs. Serdar Kuyucak, Andrew
Davies, Ray Volkas, Carmelo Pisani and Zwi Barnea.



\begin{references}
\bibitem[19]{note4} The application of such ideas to quantum gravity is an
enticing prospect, particularly since that is a natural nonlinear theory
incorporating geometric properties that $\hat{H}(\psi)$ dynamics may share.
\end{references}
\end{document}